# Measurement of dynamic interferometer baseline perturbations by means of wavelength-scanning interferometry

**Nikolai Ushakov, Leonid Liokumovich**
St. Petersburg State Polytechnical University, Polytechnicheskaya 29, St. Petersburg, Russia, 195251

**Abstract**. A novel approach for measuring fast oscillations of an absolute value of interferometer optical path difference (OPD) has been developed. The principles of frequency-scanning interferometry are utilized for registration of the interferometer spectral function, from which the OPD is calculated. The proposed approach enables one to capture the absolute baseline variations at frequencies much higher than the spectral acquisition rate. Despite the conventional approaches, associating a single baseline indication to the registered spectrum, in the proposed method a specially developed demodulation procedure is applied to the spectrum. This provides an ability to capture the baseline variations which took place during the spectrum acquisition. An analytical model describing the limitations on the parameters of the possibly registered baseline variations are formulated. The experimental verification of the proposed approach and the developed model has been performed.

**Address all correspondence to:** Nikolai Ushakov, St. Petersburg State Polytechnical University, Polytechnicheskaya 29, St. Petersburg, Russia, 195251; Tel: +7 812-552-9678; E-mail: n.ushakoff@spbstu.ru

## 1 Introduction

Fiber-optic interferometric sensors have been a subject of an extensive study of academia and industry during the last three decades[1]. Their immunity to electromagnetic radiation, low cost, small dimensions, ability to operate in harsh environments and high performance make them attractive for a great diversity of applications for measurement of temperature[2], strain[3], pressure[4], humidity[5], electric field[6] and micro-displacements[7-9] in such areas as oil and gas exploitation[4], structure health monitoring[3], nuclear energetics[10] and fundamental science[8]. The principle of the interferometric sensors is the relation of the measurable quantity $x$ (strain, temperature, etc.) with the interferometer optical path difference (OPD) via either geometrical length $L(x)$ or refractive index $n(x)$ of the media, in which the path difference is obtained.

In most of these applications it's crucial to capture the absolute value of the interferometer OPD, this can be performed by either white-light[11] techniques utilizing tunable read-out interferometer and approaches based on the registration and subsequent analysis of the interferometer spectral function. One of the most accurate spectral registering methods is wavelength-domain interferometry (WDI), demonstrating picometer-level resolutions, high absolute accuracies and large dynamic measurement range[7,9]. However, due to the limited time of spectrum measurement $T_M$ and limited spectrum acquisition repetition frequency $F_S$ ($F_S \geq 1/T_M$), the sample rate may be not enough for some applications, where higher speeds along with absolute value and high resolution are essential. A feasible solution is to use a spectrometer with higher acquisition rate. However, generally such spectrometers are based on diffraction gratings and charge-coupled devices, which cannot provide high spectral resolution and signal-to-noise-ratio (SNR). Therefore, the resolution of such sensors will be significantly reduced. In the current paper we develop an alternative technique, based on a novel signal processing approach, enabling one to overcome the disadvantage of single point per spectrum measurement and track much faster baseline fluctuations than the spectrum acquisition rate.

In the WDI techniques the optical spectral function of the interferometer is registered, which, for a two-beam interferometer (considering the low-finesse Fabry-Perot configuration with OPD=$2nL$), is given by expression below

$$S_I(L, \lambda) = S_0(L, \lambda) + S(L, \lambda), \quad (1)$$
$$S(L, \lambda) = S_M \cos[4\pi nL/\lambda + \gamma(L, \lambda)], \quad (2)$$

where $S_0 = I_1+I_2$, $S_M = 2(I_1 I_2)^{1/2}$, $I_1$ and $I_2$ are intensities of the interfering light beams; $n$ is the refractive index of the media in which the propagation takes place; $\lambda$ is the free-space light wavelength; the additional phase term $\gamma(L, \lambda)$ is induced by the phase shifts in the elements of the optical setup and a diffraction-induced phase shift if non-guided beam propagation takes place.

One of the most attractive approaches for estimation of the baseline $L$ from the registered spectrum is to approximate its variable component $S'(\lambda)$ with analytical expression (2) by means of least-squares fitting. Such fitting returns the global minimum of the residual norm, given by expression

$$R(L) = \|S'(\lambda) - S(L, \lambda)\| = \sqrt{\sum_i [S'_i - S_i(L)]^2}, \quad (3)$$

where $S'_i = S'(\lambda_i)$, $S_i(L) = S(\lambda_i, L)$, $\lambda_i = \lambda_0 + i \cdot \Delta$, $\Delta$ is the step between the spectral points, $i = -(M-1)/2,\ldots(M-1)/2$, $M$ is the number of points in digitized spectrum (for the current notation $M$ must be odd, which corresponds to the utilized interrogator and the performed simulations).

With the use of such approximation-based approach a resolution of an air-gap EFPI cavity length around 14-15 pm has been demonstrated[12]. However, the sample rate $F_S$ of conventional spectral function-registering measurements is equal to the spectrum acquisition rate, which doesn't exceed several Hz.

## 2  SIGNAL PROCESSING

In the current paper we propose a method of registering much more rapid fluctuations of the interferometer baseline. Let us consider that the wavelength-domain interferometry methods are used for interrogating the sensor, therefore, at each particular temporal moment $t_i \in [-T_M/2; T_M/2]$ (or $i$-th spectrum point) the wavelength $\lambda_i$ can be written as $\lambda_i = \lambda_0 + k_\lambda \cdot t_i$, the width of the wavelength scanning range $\Lambda = M \cdot \Delta$. It should be noted that the zero time moment $t = 0$ corresponds to $i = 0$ (the middle point of the spectrum) and the step between $t_i$ moments is equal to $1/f_D$, where $f_D$ is the sample rate of the acquired photodetector signal.

On this basis one can take into account the variation of the interferometer baseline during the spectrum acquisition time $T_M$. In this case the variable part of the interferometer spectrum $S'_I$ can be expressed as

$$S'(\lambda_i, L(t_i)) = S_M \cdot \cos\left[\frac{4\pi n L(t_i)}{\lambda_i} + \gamma[L(t_i), \lambda_i]\right] = S_M \cdot \cos\left[\frac{4\pi n L_0}{\lambda_i} + \frac{4\pi n \cdot \delta L(t_i)}{\lambda_i} + \gamma(L(t_i), \lambda_i)\right], (4)$$

where $L_0$ is the mean value of the interferometer baseline during spectrum acquisition, $\delta L(t_i)$ is the baseline variation with respect to the mean $L_0$, further denoted as $\delta L_i$ for simplicity.

For further convenience we transition from the wavelength $\lambda_i$ to the optical frequency $\nu_j = \nu_0 + k_\nu \cdot t_j$ in the interferometer spectrum expressions, as was done in, for instance[13]. It should be noted that the uniform grids of the wavelengths $\lambda_i$ and the optical frequencies $\nu_j$ do not correspond to each other, since the uniform wavelength stepping with $\Delta$ produces a non-uniform frequency grid ($\nu_i = c/n\lambda_i$), and vice versa. Hereinafter, the index $i$ will define the uniform wavelength grid, and the index $j$ will define the uniform frequency grid. On this basis the expression for the interferometer spectral function transforms to the following form

$$S'_j(L_0) = S_M \cdot \cos\left[\frac{4\pi n L_0}{c}\nu_j + \frac{4\pi n \cdot \delta L_j}{c}\nu_j + \gamma_j\right]. \tag{5}$$

The structure of the expression (5) is quite similar to the one of a quasi-harmonic signal with respect to the $\nu_j$ with carrier frequency $f_C$

$$f_C = \frac{2nL_0 k_\nu}{c} \approx \frac{2nL_0 k_\lambda}{\lambda_0^2}, \tag{6}$$

and angular modulation. It can be shown that the phase term $\gamma_j$ exhibits weak dependency on the OPD $\gamma_j(L)$ and therefore, assuming $\delta L \ll L_0$, the effect of this additional phase modulation is quite weak and the influence of the baseline variations $\delta L$ on the $\gamma_j$ can be neglected. In sequel the following notation will be used: $\gamma_j = \gamma(L_0, c/\nu_j)$, $\gamma_i = \gamma(L_0, \lambda_i)$. For the case of the constant interferometer baseline during the spectrum acquisition (5) transforms to

$$S_j(L_0) = S_M \cdot \cos\left(4\pi \cdot \frac{n}{c} L_0 \nu_j + \gamma_j\right). \tag{7}$$

This signal is quasi-harmonic since the argument increment is related not only to the equivalent frequency $2nL_0/c$, but also to the non-uniform term $\gamma_j$, which behavior must be precisely calculated or can be verified experimentally.

Even in presence of the $\delta L$ perturbation the signal $S'_j$ remains quasi-harmonic, its argument can be obtained by means of the Hilbert transform. Comparing the expressions (5) and (7), the signal processing for obtaining the $\delta L_j$ from the measured spectrum $S'_j$ can be divided into the two following steps:

1) Find the average baseline value $L'_0$ by means of approximating the measured spectrum $S'_i$ by the analytical expression (2) (the expression (7) can be applied to $S'_j$ as well, if the spectrum is measured with respect to the optical frequency). A detailed description and analysis of the approximation method used in the current study is presented in[7]. In the context of the argument demodulation task this first step is essential for finding the carrier frequency $f_C$, necessary for the following demodulation of the baseline variations.

2) By means of the Hilbert transform calculate the analytic signal for the measured spectral function $S'_j$. After that obtain the argument of the analytic signal, to which apply the standard unwrapping procedure, based on the Itoh-criterion[14]. This will produce continuous argument $\psi_j$. Then the desired difference of the arguments is calculated as

$$\Psi_j = \psi_j - 4\pi n L_0 \nu_j/c - \gamma_j, \tag{8}$$

from where the baseline variation $\delta L_j$ can be found according to the expression

$$\delta L_j = \frac{\Psi_j \cdot c}{4\pi n \nu_j}. \tag{9}$$

The use of the first step, obtaining the $L_0$ with very high accuracy enables one to find the non-perturbation part of the $\psi_j$ argument with much greater precision than detrending the $\psi_j$ and other simple methods deleting the regular components of $\psi_j$.

It should be noted that in practical optical spectrum analyzers the uniform wavelength grid is generally used, therefore, the corresponding optical frequency scale $v_i$ in (5) and (7) will be related to temporal moments as $v_i = c/(\lambda_0 + k_\lambda t_i)$, resulting in incorrect calculation of the analytical signals' phases and therefore, improper performance of the signal processing. In order to overcome this problem, two possible solutions can be proposed:
- Utilization of non-uniform fast Fourier transform (NUFFT) algorithms[15] for analytical signal calculation;
- Interpolation of the initially registered spectrum $S'(\lambda_i)$ with the uniform wavelength scale to spectrum $S'(c/v_j)$ with the uniform frequency scale before analytic signal calculation. An inverse interpolation will be needed for the calculated $\delta L_j$ signal in order to obtain the signal $\delta L_i$ (and $L_i = L'_0 + \delta L_i$), uniformly sampled with respect to time.

## 3   Method limitations

For proper performance of the proposed approach the limits on the spectrum and the amplitude of the baseline variation $\delta L_j$ must be formulated. The applicability criterion is that the spectral components of the $S'_j$ temporal representation do not decrease below zero and satisfy the Nyquist limit. For simplicity let us consider the limitations for the case of harmonic oscillation of the interferometer baseline with frequency $f_L$ and amplitude $L_m$

$$\delta L_j = L_m \cos(2\pi f_L t_j). \tag{10}$$

On the basis of expression (10), taking into account that the frequency scanning range $k_v \cdot T_M$ is much smaller than the central optical frequency $v_0$ and omitting the quasi-stationary term $\gamma_j$, the total phase of $S'_j$ can be expressed as

$$\psi(t_j) = \frac{4\pi n L_0 k_v}{c} t_j + \frac{4\pi n v_0}{c} L_m \cos(2\pi f_L t_j) = 2\pi f_C t_j + \psi_m \cos(2\pi f_L t_j). \tag{11}$$

The spectrum width of the signal with argument (11) can be estimated according to the Carson's bandwidth rule as a sum of perturbation frequency $f_L$ and frequency deviation $f_F = 4\pi n v_0 f_L L_m/c$. For adequate representation of the digitized signal, the following restrictions on the baseline variations must be fulfilled $f_L + f_F < f_C$, $f_L + f_F < f_D - f_C$. For typical values of $f_D$, $L_0$ and $k_\lambda$ ($k_v$), the second inequality is fulfilled by default, so only the first one is relevant, expressed as follows

$$f_L \cdot \left(1 + \frac{4\pi n v_0}{c} L_m\right) = f_L \cdot \left(1 + 4\pi n \frac{L_m}{\lambda_0}\right) < \frac{2n L_0 k_v}{c} \approx \frac{2n L_0 k_\lambda}{\lambda_0^2}. \tag{12}$$

For a certain device this expression serves as a relation between the maximal values of $f_L$ and $L_m$ for a signal, which can be correctly measured by the proposed approach.

On the other hand, the level of the minimal possibly detectable signal is determined by the noise level of the system. For that, the signal to noise ratio of the measured spectral function $S'_j$ must be determined as well as the relation of the noises of the initial spectrum $S'_j$ and the noises of the resulting demodulated signal $\delta L_j$.

As widely known, for an ideal phase detector, the resulting phase noise variance is related to the initial SNR of the phase modulated signal by expression[16]

$$\sigma_\varphi^2 = 0.5 \cdot SNR^{-1}, \tag{13}$$

for a SNR introduced as ratio of signal and noise powers. The sense of this relation is quite similar to the one of a Cramer-Rao bound, giving the lower limit of a sampled noisy sinusoid estimated phase's variance[17]. It should be noted that the noise power is captured in the frequency band of the photodetector (or analog to digital converter, depending on which is broader). An analogy with phase detector is applicable in our case, since the argument of the $S'_j$ signal is found by means of a Hilbert transform during the calculation of the target signal $\delta L_j$. An analytical model describing the relation of the SNR of the registered spectral function with the parameters of the optical setup was developed in[12], where a particular case of EFPI with Gaussian beam assumption was considered. As in the current paper, the WDI was analyzed, therefore, the noise mechanisms can be considered the same, resulting in two main noise contributions:
- jitter of the wavelength points during the wavelength scan, caused by the fluctuations of the signal sampling moments, characterized by a random variables $\delta\lambda_i$ (or $\delta\lambda_j$ in the above used notation) with standard deviation $\sigma_{\delta\lambda}$;
- additive noises, produced by the photo registering devices, light source intensity noises, etc., denoted as a normally distributed random term $\delta s_i$ (or $\delta s_j$), $\sigma_s=\text{stdev}\{\delta s\}$.

With this in mind, the resultant spectral function of the interferometer can be written in the following form

$$S_\text{I} = S_0 + S'_i = I_1 + I_2 + 2\sqrt{I_1 I_2}\cos\left[\frac{4\pi n(L_0+\delta L_i)}{\lambda_i+\delta\lambda_i}+\gamma(L_0,\lambda_i+\delta\lambda_i)\right]+\delta s_i. \tag{14}$$

Throughout this paper the $\gamma$ phase term will be approximated as $\gamma(L_0, \lambda_i)$, since, as shown in the section 2, its deviation due to $\delta L_i$ is quite small, and on the other hand, its dependence on the wavelength is mainly a consequence of free beam propagation, while, as shown in[12] for an extrinsic Fabry-Perot interferometer, its influence on the resultant SNR of $S'_i$ is negligible. Since the wavelength jitter and the laser intensity and photodetector noises are independent, we will consider these mechanisms separately and then will estimate their total influence on the sensor resolution.

As shown in[12] for the signal approach[7], the spectral function distortion produced by the wavelength jitter can be interpreted as an additive noise, with signal to noise ratio given by the expression

$$\text{SNR}_\text{J} = \frac{2\lambda_0^4}{(8\pi n L_0\,\sigma_{\delta\lambda})^2}. \tag{15}$$

In order to consider a most general case, we will assume that the additive noise level depends on the mean optical power incident on the photodetector. A simple yet practical approximation of this dependence by a power function will be used

$$\sigma_s = aS_0^b. \tag{16}$$

The parameters $a$ and $b$ must be obtained explicitly for a given experimental setup. On this basis, the expression for signal to noise ratio stipulated by the additive noises can be expressed as follows

$$\text{SNR}_\text{A} = \frac{S_\text{M}^2/2}{\sigma_s^2} = \frac{2I_1 I_2}{a^2\cdot(I_1+I_2)^{2b}} = \frac{V^2}{2a^2\cdot S_0^{2b-2}}, \tag{17}$$

where $V = 2(I_1 I_2)^{1/2}/(I_1 + I_2)$ is the visibility of fringes in the interferometer spectrum, $S_0 = I_1 + I_2$ is the optical power incident to the interferometer.

The resultant signal to noise ratio will be given by the expression

$$\text{SNR}_\text{T} = \left(\text{SNR}_\text{A}^{-1}+\text{SNR}_\text{J}^{-1}\right)^{-1}. \tag{18}$$

As a result, according to (9), (13) and (18) the standard deviation of the baseline measurement noise $\sigma$ can be expressed as

$$\sigma_{\mathrm{F}} = \frac{\lambda_0}{4\pi n}\sigma_{\varphi} = \frac{\lambda_0}{4\sqrt{2}\pi n}\left(\mathrm{SNR}_{\mathrm{A}}^{-1} + \mathrm{SNR}_{\mathrm{J}}^{-1}\right)^{1/2} = \frac{\lambda_0}{4\sqrt{2}\pi n}\sqrt{\frac{(8\pi n L_0\,\sigma_{\delta\lambda})^2}{\lambda_0^4} + \frac{4a^2 S_0^{2b-2}}{V^2}}. \tag{19}$$

The above expression can be used for estimating the lower bound of the detectable signal amplitude. In the form (19) it gives the standard deviation of the noises in the initial (broad) frequency band, determined by the sample rate of the photodetector and equal to $f_D/2$. However, it's more convenient to consider the noises only in the frequency band of the target signal, introducing the noise spectral density. Assuming that the resulting noise samples after the Hilbert transform demodulation are uncorrelated, one can easily find the spectral density $g_L$ of the OPD noises as

$$g_L = \sigma_{\mathrm{F}}\cdot\sqrt{\frac{2}{f_{\mathrm{D}}}} = \frac{\lambda_0}{4\pi n}\sqrt{\frac{(8\pi n L_0\,\sigma_{\delta\lambda})^2}{\lambda_0^4} + \frac{4a^2 S_0^{2b-2}}{V^2}}\cdot\sqrt{\frac{1}{2f_{\mathrm{D}}}}. \tag{20}$$

The OPD noise standard deviation $\sigma_L$ can be found as a product of the spectral density $g_L$ and square root of the target frequency bandwidth. For simplicity the equivalent carrier frequency $f_C$ (6) will be taken as the upper limit of the signal bandwidth. The resultant baseline noise in the target frequency band $[0, f_C]$ is therefore given by the expression

$$\sigma_L = g_L \cdot \sqrt{f_{\mathrm{C}}} = \frac{\lambda_0}{4\pi n}\sqrt{\frac{(8\pi n L_0\,\sigma_{\delta\lambda})^2}{\lambda_0^4} + \frac{4a^2 P_0^{2b-2}}{V^2}}\cdot\sqrt{\frac{n L_0 k_\lambda}{\lambda_0^2 f_{\mathrm{D}}}}. \tag{21}$$

Depending on a particular task, the resolution will be stipulated by different factors, however, it will be related to the $\sigma_L$ anyway.

The estimates (20) and (21) can be applied to a particular interferometric scheme if the visibility value $V$, generally dependent on the interferometer parameters, is specified. Below we will consider the case of low-finesse EFPI, which can be approximated as a two-beam interferometer. In this case the beam intensities are determined by the laser output power $P_0$, mirrors reflections and optical losses caused by divergence of a non-guided beam and a coupling coefficient[12] $\eta(L_0)$ of this beam and the fiber mode: $I_1 = P_0 \cdot R_1$, $I_2 = P_0 \cdot R_2 \cdot \eta(L_0)$. Under assumption of Gaussian profile of the fiber mode and the free beam, the coupling coefficient and the fringe visibility are given by the expressions

$$\eta = \frac{(\pi n w_0^2)^2}{L^2\lambda^2 + (\pi n w_0^2)^2}, \qquad V = \frac{2\sqrt{R_1 R_2}}{R_1 \cdot\left[(L\lambda/\pi n w_0^2)^2 + 1\right]^{1/2} + R_2}, \tag{22}$$

where $w_0$ is the fiber mode field radius. In such a manner, the standard deviation of the OPD noise for EFPI is

$$\sigma_L = \frac{\lambda_0}{4\pi n}\sqrt{\frac{f_{\mathrm{C}}}{\mathrm{SNR}_{\mathrm{T}}\cdot f_{\mathrm{D}}}} = \frac{\lambda_0}{4\pi n}\sqrt{\frac{(8\pi n L_0\,\sigma_{\delta\lambda})^2}{\lambda_0^4} + \frac{a^2(R_1 + R_2\eta(L_0))^{2b}}{P_0^{2-2b} R_1 R_2 \eta(L_0)}}\cdot\sqrt{\frac{n L_0 k_\lambda}{\lambda_0^2 f_{\mathrm{D}}}}, \tag{23}$$

where the expression for the coupling coefficient (22) wasn't substituted in order to avoid an excessive bulkiness.

The lower bound of the perturbation frequency is the simplest and is formulated as follows. For proper performance of the spectrum approximation approach[7], applied at the first step of the processing, the linear component of the baseline perturbation must be minimized, therefore, the period of the baseline oscillation must be less or equal to the spectrum measurement time $T_\mathrm{M}$.

The rapid slope of the baseline during the spectrum measurement time results in an abrupt approximation error. An approach overcoming this issue was proposed and implemented in[18].

In order to achieve the resolutions given by the expressions (21) and (23), the calculated baseline variation with uniform temporal sampling $\delta L_i$ must be filtered by a low-pass filter with cut-off frequency $f_C$. By doing so, the effort in displacement resolution will be $\sim(f_D/2f_C)^{-1/2}$ for the case of white noise, which, for the practical frequencies (see section 4) is greater than one order. Recall the approximating baseline estimation approach, it corresponds to the filtering performed with the lowest physically meaningful cut-off frequency $f_C = 1/T_M$. Let us note the correspondence between the obtained estimations (21), (23) and the quasi-static case, returning a single point for a measured spectral function (as done in[7,9,12,13]): considering an extreme situation $f_C = 1/T_M$ and substituting it and $f_D = M/T_M$ into (23), one obtains

$$\sigma_L = \frac{\lambda_0}{4\pi n}\sqrt{\frac{1}{\mathrm{SNR}_T \cdot M}} = \frac{\lambda_0}{4\pi n} \cdot \sigma_{\varphi\mathrm{CRLB}} = \sigma_{L\,\mathrm{CRLB}}, \qquad (24)$$

where $\sigma_{\varphi\mathrm{CRLB}}$ is a Cramer-Rao bound[17] on the standard deviation of the estimate of a noisy sinusoid's phase; $\sigma_{L\,\mathrm{CRLB}}$ – Cramer-Rao bound recalculated to the interferometer baseline domain. Substituting the parameters of the practical setup (see section 4), one obtains the dependency $\sigma_L \approx 0.9 \cdot 10^{-3} \cdot \mathrm{SNR}_T^{-1/2}$, which is in a good accordance with noise influence estimation on the resolution of the approximation-based baseline estimation approach, obtained in[12], $\sigma_L = 1.1 \cdot 10^{-3} \cdot \mathrm{SNR}^{-1/2}$. Therefore, the conventional approximation-based techniques are a special case of the proposed one in the limit $f_C = 1/T_M$.

## 4 Experimental demonstration

The proposed approach was implemented and tested experimentally. Spectra measurements were performed using the optical sensor interrogator National Instruments PXIe 4844, installed on PXI chassis PXIe 1065, controlled by PXIe 8106 controller. Spectrometer parameters are the following: scanning range 1510÷1590 nm (spectral interval width $\Lambda = 80$ nm, mean wavelength $\lambda_0 = 1.55$ µm), spectral step $\Delta = 4$ pm (number of spectral points $M = 20001$), scanning speed $k_\lambda = 2.4$ µm/s, spectrum acquisition time $T_M \approx 0.035$ s, output power $P \approx 0.06$ mW, wavelength scanning speed $k_\lambda = 2.4$ µm/s, photodetector sample rate $f_D = 570$ kHz, spectral points jitter $\sigma_{\delta\lambda} = 1$ pm. The parameters $a$ and $b$ for the current interrogator were found experimentally in[12] and have the following values: $a = 8.47 \cdot 10^{-4}$, $b = 0.81$. The experimental setup is schematically illustrated in figure 1.

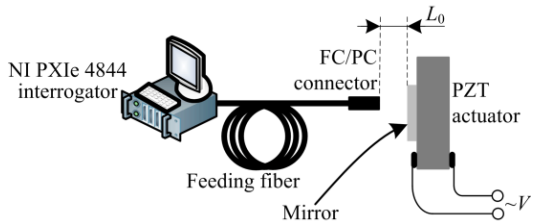

**Fig. 1.** Experimental setup.

Examined interferometer was formed by the face of the SMF-28 fiber (reflectivity at the air-fused silica bound $R_1 = 3.5\%$) with mode field radius $w_0 = 5.2$ µm, packaged in FC/PC connector and an external mirror with $R_2 \approx 90\%$, adjusted to the PZT actuator. The controlling voltage for the PZT was generated by the PXIe 5421 signal generator, installed on the same PXI chassis.

The efficiency of the utilized PZT actuator was approximately 100 nm/V in the frequency range 10 – 1000 Hz. The mean baseline value $L_0$ was varied within the interval 550 – 750 μm, resulting in the $S'_i$ carrier frequency $f_{Sc} \approx 1100 - 1500$ Hz.

For the setup parameters mentioned above, relations of the amplitude $L_m$ and frequency $f_L$ limits for different baseline values $L_0$ are illustrated in figure 2. In the diagram in figure 2, the lower limit on the detectable amplitude was assumed $\sqrt{2}\cdot\sigma_L$, estimated according to (23) with the resulting SNR=1. This condition can differ for different tasks and further signal processing. In such a manner, the final applicability limits will be determined by intersection of the criteria (12), (23) and $f_L>1/T_M$.

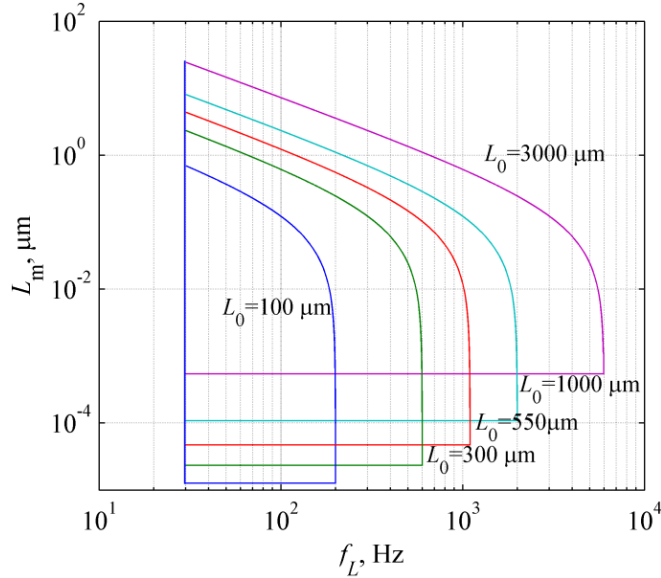

**Fig. 2.** Limits of the amplitude and frequency of the harmonic baseline variation for different baseline values.

In the performed experiments, described below, a voltage with the following parameters was applied to the PZT:
- the excitation voltage amplitude was varied from 0.05 to 3 V, resulting in the EFPI baseline variation amplitudes from 5 nm to 0.3 μm;
- the frequency was varied from 50 to 1000 Hz;
- the oscillation shape was either harmonic or triangular.

The measured signals for the case of harmonic baseline oscillations with mean baseline value $L_0 \approx 550$μm are shown in figure 3. Three cases are demonstrated: frequency 200 Hz, amplitude 90 nm; frequency 200 Hz, amplitude 7 nm; frequency 800 Hz, amplitude 12 nm.

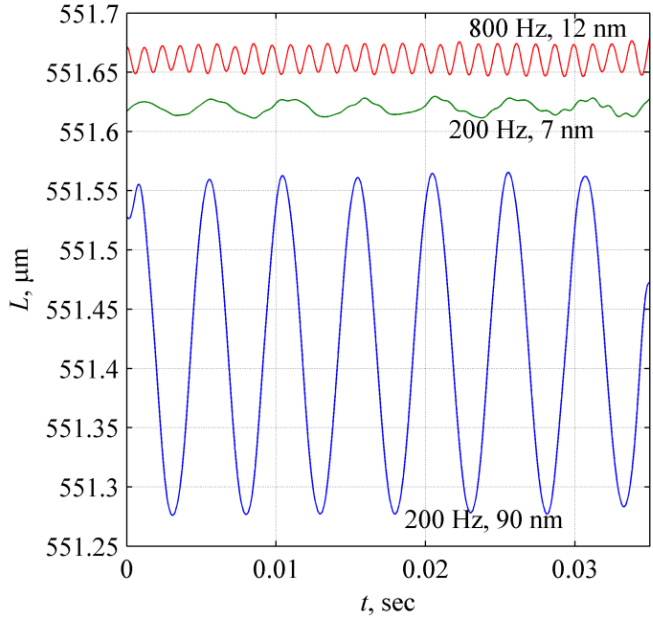

**Fig. 3.** Measured baseline variations.

In figure 4 the spectra of the above-shown signals are demonstrated. The attained noise spectral density, calculated as median level of the signal spectra in the frequency band $[0, f_C]$ was $g'_L = 1.1$ pm/Hz$^{1/2}$, which is in a good correspondence with the value estimated according to the expression (20) $g_L = 1.9$ pm/Hz$^{1/2}$. Considering the frequency band $[0, 1/T_M]$, corresponding to the approximation-based approach[7,12], the level of captured noises will be $g_L/\sqrt{T_M} \sim 10.2$ pm, which is in accordance with the observed noise-influenced value, reported in[12].

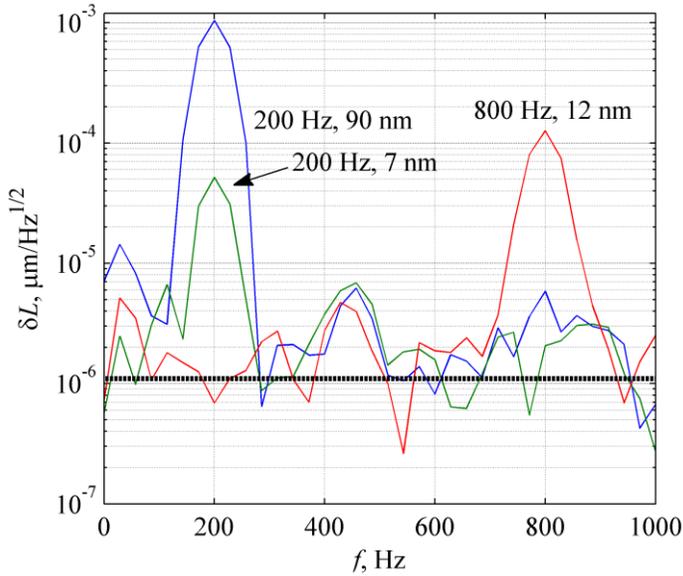

**Fig. 4.** Spectral density of the measured baseline oscillations for $L_0 \approx 550$ μm (solid lines), noise spectral density estimated according to (20) (dotted line).

The experimental signals for the other parameters of the perturbation (triangular variations of the baseline) and the setup (other mean baseline values) can be found in[19] and the corresponding results are also in a good correspondence with the predictions of the developed theoretical model.

## 5   Conclusions

In the current paper a novel approach is proposed, enabling one to overcome the conventional limitations of the wavelength-domain absolute interferometry, where a single baseline value is obtained according to a single acquired interferometer spectrum. Instead, using the proposed signal processing approach, one is able to track the fast deviations of the absolute baseline value that take place during the spectrum acquisition. The upper limits on the frequency and amplitude of the perturbation are stipulated by the no-aliasing conditions and are related to the mean baseline value $L_0$. The lower limit on the signal frequency is determined by the spectrum measurement time $T_M$ and inquires that the perturbation frequency $f_L>1/T_M$. The lower limit on the signal amplitude is determined by the signal to noise ratio of the measured interferometer spectrum, an analytical model relating the baseline resolution with the optical setup parameters was established. An experimental demonstration of the proposed baseline measurement approach with the extrinsic Fabry-Perot interferometer was performed, the attained resolution is in a good agreement with the estimations performed with the developed model.